\documentclass[footinbib,PRB,twocolumn,amsmath,amssymb,floatfix,showpacs]{revtex4-1}
\usepackage{epsfig,psfrag,subfigure}
\usepackage{colordvi}

\def\bm#1{\mbox{\boldmath$#1$}}

\def\nopc{Q}
\def\nop{{\rm{\mathbf\nopc}}}

\def\ldgthree{v}
\def\ldgfour{w}

\def\taut{t_R}

\def\mycorrel{{\cal C}}
\def\correllength{\xi}

\def\hdelta{{\mathchar'26\mkern-9mu\delta}}
\newcommand{\be}{\begin{equation}}
\newcommand{\ee}{\end{equation}}
\newcommand{\ba}{\begin{eqnarray}}
\newcommand{\ea}{\end{eqnarray}}
\newcommand{\bw}{\begin{widetext}}
\newcommand{\ew}{\end{widetext}}
\newcommand{\Tr}{{\rm{Tr}\,}}
\newcommand{\rv}{{\bm{r}}}
\newcommand{\xv}{{\bm{x}}}

\newcommand{\zv}{{\bm{z}}}
\newcommand{\pv}{{\bm{p}}}

\begin{document}

\title{Polydomain structure and its origins in isotropic-genesis nematic elastomers}

\author{Bing-Sui~Lu$^{1}$}
\author{Fangfu Ye$^{1}$}
\author{Xiangjun Xing$^{2}$}
\author{Paul M.~Goldbart$^{1}$}
\affiliation{%
$^1$Department of Physics and Institute for Condensed Matter Theory,
University of Illinois at Urbana-Champaign, 1110 West Green Street, Urbana, IL 61801\\
$^2$Institute of Natural Sciences and Department of Physics,
Shanghai Jiao Tong University, Shanghai 200240 China
}

\date{\today}

\begin{abstract}
We address the physics of nematic liquid crystalline elastomers randomly crosslinked in the isotropic state. To do this, we construct a phenomenological effective replica Hamiltonian in terms of two order-parameter fields: one for the vulcanization, the other for nematic alignment.   Using a Gaussian variational approach, we analyze both thermal and quenched fluctuations of the local nematic order, and find that, even for low temperatures, the macroscopically isotropic polydomain state is stabilized by the network heterogeneity.  For sufficiently strong disorder and low enough temperature, our theory predicts unusual, short-range oscillatory structure in (i.e., anti-alignment of) the local nematic order.  The present approach, which naturally takes into account the compliant, thermally fluctuating and heterogeneous features of elastomeric networks, can also be applied to other types of randomly crosslinked solids.
\end{abstract}


\maketitle

\section{Introduction}
\label{sec:intro}
Liquid crystal elastomers~\cite{LCE:WT} are fascinating materials, in which the liquid crystalline order is strongly coupled to the elasticity of the underlying elastomeric network.  This gives rise to
novel properties, such as soft~\cite{olmsted,golubovic} and semi-soft elasticity~\cite{verwey,ye}, which is of interest in fundamental research and, furthermore, has significant potential for technological applications.  From the perspective of statistical physics, these materials, like any other randomly crosslinked systems, pose important conceptual challenges, as their physical properties do not depend just on the conditions under which various physical quantities are measured.  They also depend on both the conditions under which the system is prepared and the network heterogeneities necessarily generated via the random crosslinking preparation process.  A coherent theory that takes all these factors into account requires two statistical ensembles, i. e. {\em the preparation ensemble and the measurement ensemble}.  This essential point was qualitatively understood by Deam and Edwards and by de~Gennes as early as the 1970's~(see, e.g., Refs.~\cite{Deam-Edwards-1976,polymer:deGennes}).
Its full significance however has not been explored previously.

One such challenge concerns the origins, structure and elasticity of the polydomain state in isotropic-genesis nematic elastomers (IGNEs; i.e., elastomers that were crosslinked in the isotropic state)~\cite{elias,uchida-2d,uchida,selinger,terentjev1,terentjev2,terentjev3,feio,urayama,biggins-short,biggins-long}.  It has always been observed experimentally that IGNEs at low temperature are characterized by a polydomain (PD) state, i.e., a macroscopically isotropic state consisting of well-defined domains of randomly oriented nematic order.  The typical domain-size is of the order of microns.  Furthermore, PD IGNEs exhibit much softer stress-strain curves than PD nematic elastomers crosslinked in nematic state, as demonstrated by recent experiments of Urayama's group \cite{urayama}.

A heuristic argument, of the type first developed by Imry and Ma~\cite{ImryMa} for systems at low temperature having quenched random fields with short-range correlations, predicts that long-range order involving continuous symmetry breaking is precluded in systems of spatial dimensions fewer than four.  This argument surely applies to nematogens in the presence of an immobile random substrate, such as aerogel~\cite{radzihovsky,feldman}.  However, for nematic elastomers, the network elasticity induces a long-range interaction between nematic directors.  Furthermore, it is not a priori clear whether the effects of network heterogeneities can be properly modeled as quenched random fields.  Hence, the original Imry-Ma type of argument does not necessarily apply.  Nevertheless, absent any convincing alternative, this traditional random-field approach has been adopted by various authors~\cite{uchida-2d,uchida,selinger,terentjev1,terentjev2,terentjev3}.

Simulation work has been done by Uchida~\cite{uchida, uchida-2d} on a simple, two-dimensional, random-field model of nematic elastomers.  He found that IGNEs always exhibit the PD state at a sufficiently low temperature. On the other hand, Terentjev et al.~\cite{terentjev1, terentjev2,terentjev3} have used Gaussian variational methods to study a similar model, and they concluded that replica symmetry is broken, indicative of glassy low-temperature properties in PD IGNEs.  Xing and Radzihovsky~\cite{Xing-Radz2,Xing-Radz3} start from a disorder-free model of a MD nematic elastomer, turn on an infinitesimal quenched random field and study the consequent stability of the long-range nematic order, via renormalization-group techniques.  They found that nematic order persists even below three spatial dimensions.   The random-field idea is also implicit in the more recent works by Biggins et al.~\cite{biggins-short,biggins-long}, which employs a quasi-convexification method to construct textured, soft deformations of IGNEs, and thus allows these authors to obtain sharp bounds on the stress-strain curve.

The aim of the present paper is to explore the nature of quenched disorder in IGNEs, and its impact on local nematic order, within the framework afforded by vulcanization theory~\cite{Vulcan_Goldbart,peng,Zippelius-EPJB-2002,XPMGZ-NE-VT}.  We find that the most relevant type of disorder does indeed resemble a quenched random field, but with an important difference which we shall discuss in detail below.  By using the replica Gaussian variational approach, we calculate the thermal as well as sample-to-sample fluctuations of the local nematic order.  Our principal findings are that the PD state is always stable, even at sufficiently low temperature and, furthermore, that for sufficiently strong disorder, the local frozen nematic order has an oscillatory character (i.e., anti-alignment) at short distances.

\section{Order parameters, free energies and correlators}
\label{sec:Landau}
\begin{figure}
	\centering
		\includegraphics[width=0.4\textwidth]{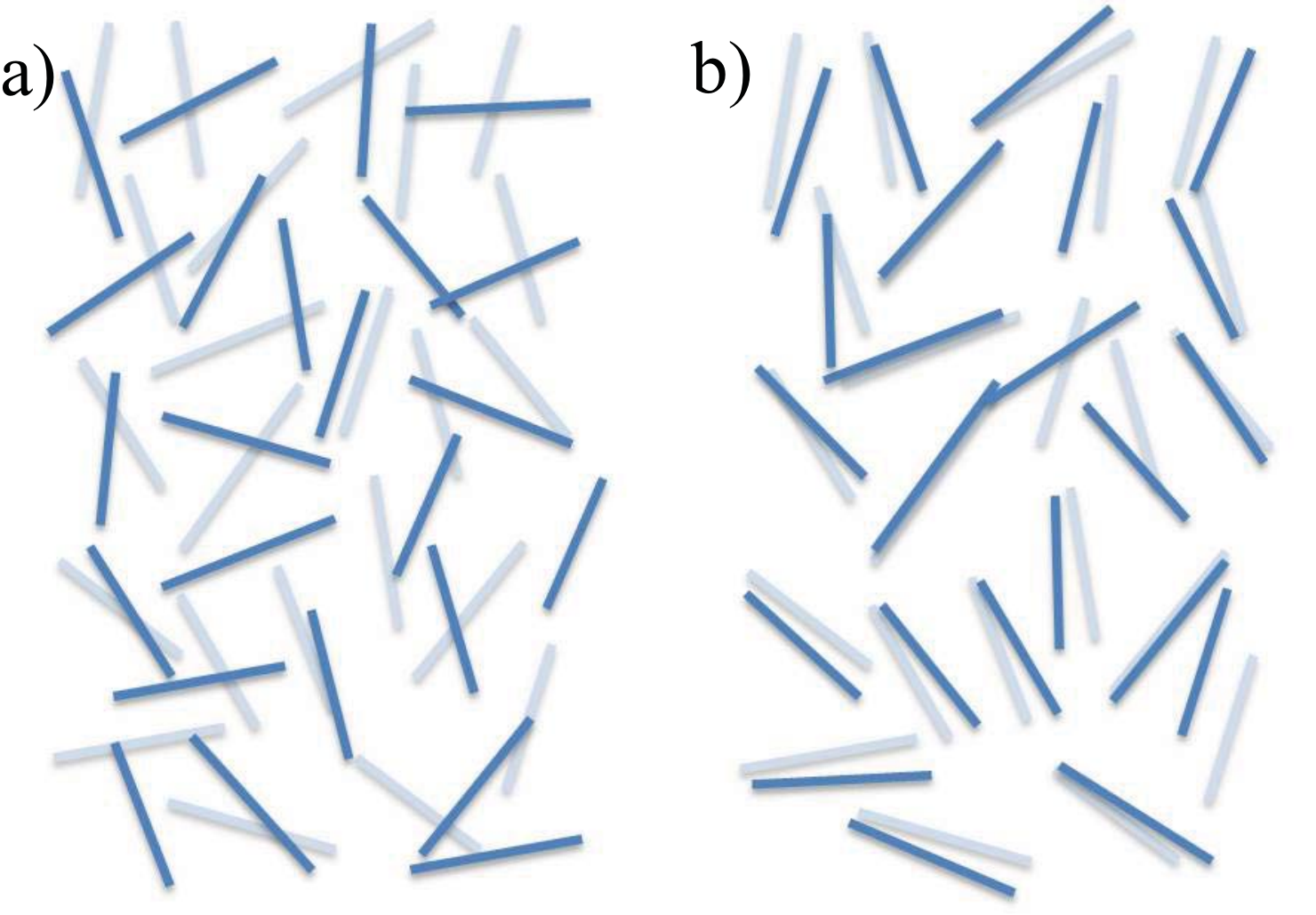}
\caption{(Schematic) snapshots of nematogen orientations at a particular instant (blue unshaded),
and at a much earlier instant (shaded).
(a)~A conventional liquid crystal in the isotropic state.
Such systems do not memorize the local pattern of nematogen alignment indefinitely.
There is no correlation between the orientations of blue and shaded nematogens that are depicted near one another.
(b)~A liquid cystalline ealstomer in the macroscopically isotropic state.
Such systems do memorize the local pattern of nematogen alignment indefinitely.
The orientations of blue and shaded nematogens depicted near one another are  correlated. }
  \label{fig:trappingin}
\end{figure}

The physics of randomly crosslinked systems depends on both the state under which the system is crosslinked, and the state under which it is measured.   It is therefore of crucial importance to distinguish between these two states.  We call the former the {\em preparation state} and the latter the {\em measurement state}.  The local nematic order $\nop^0$ in the preparation state and its counterpart $\nop$ in the measurement state are, in principle, distinct, and this distinction is naturally captured by the formalism of vulcanization theory.  For IGNEs crosslinked at high temperature, $\nop^0$ is negligibly small.  In contrast, although the macroscopic average of $\nop$ vanishes in IGNEs, its local value can be substantial; this would reflect random local frozen nematic alignment.  It is also important to note the fundamental asymmetry  in the \textit{causal\/} ordering of $\nop^0$ and $\nop$: whilst $\nop^0$ can influence $\nop$, the converse is forbidden.  In the framework of vulcanization theory, this causality is naturally enforced via the replica trick, in which the measurement-state order parameter is replicated $n$ times, $\{\nop^{\alpha}; \alpha = 1,\ldots, n\}$, and the number of copies $n$ is later taken to zero.

As IGNEs are quench-disordered systems, we need to distinguish between thermal averaging (i.e., averaging over thermal fluctuations), denoted $\langle \cdots \rangle$, and disorder averaging (i.e., averaging over the realizations of the crosslinking), denoted $[ \cdots ]$.  Assuming that the system is self-averaging, it follows that every experimentally measured quantity has already been disorder-averaged.  Various disorder- and thermal-averaged quantities involving the local nematic order in the measurement state $\nop$ can be constructed.   To lowest order in $\nop$, one can construct the quantity $[\langle Q(\rv) \rangle ]$; however, this quantity trivially vanishes, owing to the macroscopic isotropy of the PD state.  At second order in $\nop$, one can form the following two, distinct, averaged quantities: (i) the \textit{thermal correlator}
\begin{subequations}
\begin{equation}
\label{eq:thermcorrdef}
\mycorrel^{T}({\bm r}-{\bm r}^{\prime})
:=
\Tr\big[\big\langle
\big(\nop({\bm r})-\langle\nop({\bm r})\rangle\big)\,
\big(\nop({\bm r}^{\prime})
-\langle\nop({\bm r}^{\prime})\rangle\big)
\big\rangle\big],
\end{equation}
and (ii) the \textit{glassy correlator}
\label{eq:coredefs}
\begin{equation}
\label{eq:glasscorrdef}
\mycorrel^{G}({\bm r}-{\bm r}^{\prime})
:=
\Tr[\langle\nop({\bm r})\rangle\,
  \langle\nop({\bm r}^{\prime})\rangle].
\end{equation}
\end{subequations}
Although we are focusing on the Cartesian scalar aspects of the correlators, it is straightforward to reconstruct the full structure of the corresponding fourth-rank Cartesian tensors, by appending suitable isotropic tensor factors constructed from Kronecker deltas.

The glassy correlator automatically vanishes for isotropic liquids.  For IGNEs, however, it is nonzero and encodes the correlation in the randomly frozen nematic pattern.   The difference between these two states is qualitatively indicated in Fig.~\ref{fig:trappingin}.  The value if the glassy correlator at the origin, $\mycorrel^{G}(\rv)|_{\rv=\bm{0}}$, is the nematic analogue of the Edwards-Anderson order parameter for spin glasses~\cite{Edwards-Anderson-SpinGlass-1975}, and measures the magnitude of the local frozen nematic order.  We shall call this the \textit{glassy order parameter}.   On the other hand, the thermal correlator can be used to detect the existence of a continuous phase transition.  If such a transition exists, the thermal correlator would diverge at the critical temperature.  If a discontinuous transition occurs, however, the correlator would diverge at the spinodal point.

To address the interplay between nematic ordering and network heterogeneity in randomly crosslinked systems, we shall also need the vulcanization order parameter $\Omega({\bm r}^{0},{\bm r}^{1},\ldots,{\bm r}^{n})$, which is a scalar function of $1+n$ replicated three-dimensional position vectors $\{\rv^{\alpha}; \alpha = 0, 1, \ldots, n\}$ (see, e.g., Refs.~\cite{Vulcan_Goldbart,peng,Zippelius-EPJB-2002,XPMGZ-NE-VT}).
Up to an additive constant, it gives the joint probability density function (pdf) for the results of the following $1+n$ independent measurements: At a time just prior to crosslinking, a monomer is found at $\rv^0$, whilst in $n$ independent measurements made long after crosslinking, the same monomer is found at $\rv^1,\ldots, \rv^n$.  This order parameter detects and characterizes the positional localization of monomers that emerges at the gelation/vulcanization transition.  In the liquid phase, all monomers are delocalized, so the joint pdf is a trivial constant, and hence the order parameter $\Omega$ vanishes identically.  In the gel/rubber phase, a fraction $G$ of monomers are localized.  Their positions in the $1 + n$ distinct measurements are then correlated with one another, and consequently the order parameter has a nontrivial, mean-field profile, given by
\be
\label{eq:barredOP}
\bar{\Omega} (\hat{r}) =
G\!\! \int\!\!  d^3 z \!\!
\int\!\!  d \tau p(\tau) \left(\frac{\tau}{2\pi}\right)^{\tfrac{3}{2}(n+1)}\!\! \!\!
e^{
- \frac{\tau}{2}
\sum_{\alpha = 0}^n
|\rv^{\alpha} - \zv|^2}
 - \frac{G}{V^{n}},
\ee
where $\hat{r}:= (\rv^0,\ldots, \rv^{n})$ and $V$ is the volume of the system.
Note that the saddle point order parameter $\bar{\Omega}$ is invariant under simultaneous translation of all replicated vectors $\hat{r}:= (\rv^0,\ldots, \rv^{n})$.
The parameter $\tau$ has the physical meaning of the local inverse square localization length for the monomers, and it fluctuates from monomer to monomer with pdf $p(\tau)$.
In this work, we shall make the simplifying assumption that the localization length has a sharp value $\xi_L$ rather than being distributed, i.e.,
$p(\tau) = \delta (\tau - \xi_L^{-2})$.
This does not change our essential results.

It is rather significant that the heterogeneous nature of elastomers is captured already at the mean field level of vulcanization theory.   At long lengthscales, the most important
coupling between the vulcanization order parameter $\bar{\Omega}$ at the saddle point and the nematic order parameters $\nop^{\alpha}$ reads
\begin{eqnarray}
\label{coupling}
- h\!\!\int d \hat{r}
 \sum_{\alpha,\beta=0}^{n}\!\!\!
\nabla_i^{\alpha}
\nabla_j^{\alpha}
\bar{\Omega}(\hat{r})
\nabla_k^{\beta}
\nabla_l^{\beta}
\bar{\Omega}(\hat{r})
\nopc_{ij}^{\alpha}({\bm r}^{\alpha})
\nopc_{kl}^{\beta}({\bm r}^{\beta}),
\end{eqnarray}
where $d\hat{r}$ indicates multiple integration over all replicated positions $\rv^\alpha$, and a summation is implied over the Cartesian coordinates $i,j,\ldots$.   In the above summation, the replica-diagonal terms and the replica-off-diagonal terms contribute qualitatively distinctly from one another.  In particular, the replica-diagonal terms can be omitted as they simply renormalize the quadratic term in the Landau-de~Gennes free energy~\cite{footnote-one-replica}.
We proceed to integrate out the coordinates
${\bm r}^{\gamma}$ (with $\gamma\neq\alpha,\beta$).
Hence, Eq.~(\ref{coupling}) is transformed into the following, essentially nonlocal, quadratic term:
\begin{subequations}
\label{coupling-0}
\begin{eqnarray}
\label{coupling-1}
&&
-\frac{h}{2}\!\int\! d^3 r\, d^3 r^{\prime}\,
\omega({\bm r}-{\bm r}^{\prime})\!\!
\sum_{\alpha,\beta=0\atop{(\alpha\neq\beta)}}^{n}
\!\!\Tr
\nop^{\alpha}({\bm r})\,
\nop^{\beta}({\bm r}^{\prime}),
\\
&&
\sqrt{2}\omega(\rv):=
(4\pi/\correllength_L^2)^{7/2}
G^2\,
e^{-\vert\rv\vert^{2}/ 2\xi_L^2},
\end{eqnarray}
\end{subequations}
where $G$ is the gel fraction, and $\omega$ is a kernel that represents the disordering effects of the random network on the nematic order.  The characteristic lengthscale beyond which $\omega(\rv)$ is suppressed is $\correllength_L$.  This suppression reflects the liquid nature of the network at short lengthscales but not at lengthscales longer than $\correllength_{L}$.

 For IGNEs crosslinked at a very high temperature, the local nematic order $\nop^0$ in the preparation state is negligibly small and can therefore be ignored.
The full free energy of IGNEs is a combination of the Landau-de~Gennes free energy for local nematic order in the measurement state and the coupling term, Eqs.~(\ref{coupling-0}):
\begin{eqnarray}
\label{H_eff}
H\left[\nop \right]
&=&
\!\int d^{3}{r}\,
\sum_{\alpha=1}^{n}
\bigg\{
 \frac{T_r}{2}\Tr\nop^{\alpha}\nop^{\alpha}\,
+\frac{K}{2} \Tr
\nabla_i\nop^{\alpha}
\nabla_i\nop^{\alpha}
\nonumber\\
&-&\!\frac{\ldgthree}{3}
\Tr\,(\nop^{\alpha})^3
+\frac{\ldgfour}{4}
({\rm Tr}\,\nop^{\alpha}
\nop^{\alpha})^2\,
\bigg\}
\\
&-&\!\frac{h}{2}\!\!
\sum_{\alpha,\beta=1\atop{(\alpha\neq\beta})}^{n}\!\!
\int \!\! d^3 r\,d^3 r^{\prime}
\omega({\bm r}-{\bm r}^{\prime})
\Tr\,
\nop^{\alpha}({\bm r})\,
\nop^{\beta}({\bm r}^{\prime}).
\nonumber
\end{eqnarray}
Here, $T_r$ is the reduced temperature in the measurement state.  In addition, $K$ is proportional to the Frank constant in the one-constant approximation.  It is important to note that any standard replica treatment of a phenomenological theory in which the nematic order is coupled linearly to a quenched Gaussian random field with correlator $\omega(\rv)$ would necessarily lead, in addition to the replica off-diagonal term in Eq.~(\ref{H_eff}), to a nonvanishing replica-diagonal term having precisely the same kernel as its off-diagonal counterpart.  This absence of the replica-diagonal term is a salient feature of vulcanization theory.  {\em It demonstrates that the quenched disorders in elastomers can not be simply modeled as random field}.  As we shall show below, it is also responsible for the oscillatory spatial structure for IGNEs in the strong-disorder regime.  Standard random-field models would not yield such oscillatory structure.

Focusing on the quadratic parts of the effective Hamiltonian~(\ref{H_eff}), we can obtain the bare thermal ($\mycorrel^T_0$) and glassy ($\mycorrel^G_0$) correlators [see Eqs.~(\ref{eq:coredefs})] of the nematic order parameter in wave-vector space:
\begin{subequations}
\label{corr-bare}
\ba
\label{corr-bare-1}
\mycorrel^T_0(\pv) &=&
\frac{R_l }{t + \kappa^2 + \Delta e^{-\kappa^2/2}},
\\
\mycorrel^G_0(\pv) &=&
\frac{R_l \Delta e^{-\kappa^2/2}}
{\left( t + \kappa^2 + \Delta e^{-\kappa^2/2}\right)^2}.
\label{corr-bare-2}
\ea
\end{subequations}
Here, $R_l := \xi_L^2/K$,
whilst $t:= T_r \xi_L^2/K$,
$\Delta := 32{\pi}^3 hG^2/K \xi_L^2>0$ and
$\kappa := p \xi_L$ are the dimensionless rescaled measurement-temperature, disorder strength, and wavevector.
The appearance of $\Delta$ in the denominators, in combination with the {\it scale-dependent\/} factor $\exp({-\kappa^2/2})$, is a significant result of the present approach.  It follows directly from the absence of the replica-diagonal contributions to the term~(\ref{coupling-0}), and therefore cannot result from a standard random-field approach.  Originating in the random network, and proportional to the shear modulus of the network~\cite{footnote:shearmodulus}, it leads to a scale-dependent, {\it downward\/} renormalization of the reduced temperature.

The bare correlators exhibit instability at some negative reduced temperature $t<0$.  For $\Delta < 2$, i.e., at weaker disorder, the denominator factor
$t + \kappa^2 + \Delta e^{-\kappa^2/2}$ has a minimum at $\kappa = 0$.  Consequently, both of the correlators given in Eqs.~(\ref{corr-bare}) diverge, at $\kappa = 0$, at a critical (reduced, rescaled) temperature $-\Delta$.
This suggests a continuous transition to a uniformly ordered (i.e., macroscopically anisotropic) nematic state.
By contrast, for $\Delta > 2$, i.e., at stronger disorder, the denominator factors have a minimum at
$\kappa^2 = 2\ln (\Delta/2)$;
therefore, both bare correlators exhibit nonzero-wavelength divergences at a (reduced, rescaled) temperature
$-2\ln(e \Delta /2)$.
This suggests a continuous phase transition to a state having periodic modulations.  As we shall see shortly, however, neither of these putative transitions actually occurs, both of them being precluded by the quenched randomness.

\section{Gaussian variational approach to thermal and glassy correlators}
\label{sec:diagnostics}

To investigate the fate of the system at temperatures below these putative  instabilities, we employ the Gaussian variational method (see, e.g., Ref.~\cite{feynman}) for systems having quenched randomness~\cite{Mezard-Parisi-replica-variational-1990,Mezard-Parisi-replica-variational-1991,shakhnovich,dobrynin}.  We choose the corresponding quadratic trial Hamiltonian $H_0$ to be
\ba
&&H_{0}=
\frac{1}{2}\sum^{n}_{\alpha,\beta=1}
\int_{\pv_1,\pv_2}\,
\Gamma^{\alpha\beta} (\pv_1,\pv_2)\times
\nonumber\\
&&\qquad \Tr
\big(\nop^{\alpha}(\pv_1)-{\bar{\nop}}(\pv_1)\big)
\big(\nop^{\beta}(\pv_2) -{\bar{\nop}}(\pv_2)\big),
\ea
which generically features a nonvanishing mean nematic order parameter ${\bar{\nop}}$ together with a kernel $\Gamma^{\alpha\beta}$~\cite{footnote:kernel}.
Here, the notation $\int_{\pv}$ denotes $\int d^3 p /(2\pi)^3$.
Assuming that replica permutation symmetry remains intact~\cite{footnote:rsb}, we may parametrize the replica-space inverse of $\Gamma^{\alpha\beta}$ as
\begin{equation}
\big(\Gamma^{-1}\big)^{\alpha\beta}(\pv_1,\pv_2):=
\hdelta(12)\big(\mycorrel^T(\pv_1) \delta^{\alpha\beta}
+\mycorrel^G(\pv_1)\big),
\end{equation}
where $\hdelta(12\cdots m)$ denotes $(2\pi)^3\delta(\pv_1+\pv_2+\cdots+\pv_m)$, and $\mycorrel^T(\pv)$ and $\mycorrel^G(\pv)$ are the {\it renormalized\/} thermal and glassy correlators, in which fluctuations are approximately accounted for.  Both of them, together with $\bar{\nop}$, are to be determined self-consistently, by minimizing the resulting variational free energy
\begin{equation}
F_{\rm var}:=\big\langle{H-H_{0}}\big\rangle_{H_{0}}-\ln
\int \prod_{\alpha=1}^{n}{\cal D}\nop^{\alpha}\,{\rm e}^{-H_{0}},
\label{F_var-def}
\end{equation}
where $\langle\cdots\rangle_{H_{0}}$ denotes averaging with respect to the Boltzmann weight $e^{-H_{0}}$.  The explicit form of $F_{\rm var}$ is
\ba
\label{F_var}
\frac{2}{5nV}F_{\rm var}&=&\frac{1}{R_l}\!\!\int_{\pv}\!\! \big( t + \kappa^2+
\Delta e^{-\kappa^2/2} \big) \big( \mycorrel^G + \Tr \bar{\nop}^2/5V\big)\nonumber\\
&+&
\frac{1}{R_l}\!\!\int_{\pv}\!\!\big( t + \kappa^2 \big) \mycorrel^T
+ \frac{7w}{5{V}}\!\!\int_{\pv_1}\!\!\!\Tr \bar{\nop}^2\!\! \int_{\pv_2}\!\!\! (\mycorrel^T + \mycorrel^G)
\nonumber\\
&+&
\frac{7w}{2}\left( \int_{\pv}(\mycorrel^T+ \mycorrel^G) \right)^2
- \int_{\pv} \left( \ln \mycorrel^T + \frac{ \mycorrel^G}{ \mycorrel^T} \right)
\nonumber\\
&+&
\frac{2\ldgthree}{15{V}}
\int_{\pv_1,\pv_2,\pv_3}\!\!\!\!\!\!
\hdelta(123)\Tr \bar{\nop}^3\nonumber\\
&+&
\frac{w}{10{V}}
\int_{\pv_1,\pv_2,\pv_3,\pv_4}\!\!\!\!\!\!\!\!
\hdelta(1234)
\big(\Tr \bar{\nop}^2\big)^2.
\ea
By minimizing $F_{\rm var}$ with respect to $\mycorrel^T$ and $\mycorrel^G$
for a given value of $\bar{\nop}$, and defining a renormalized reduced temperature $\taut$ via
\begin{equation}
\taut= t + 7  \thinspace\ldgfour R_l
\int_p
\left(
\mycorrel^{T}
+\mycorrel^{G}
+\frac{1}{5V} \Tr\bar{\nop}^2
\right),
\label{relation-tau-t}
\end{equation}
we obtain for the {\it renormalized\/} thermal and glassy correlators the results
\begin{subequations}
\label{corr}
\begin{eqnarray}
\label{corr-1}
\mycorrel^{T}(\pv)
&=&
\frac{R_l}{\taut + \kappa^2 +  \Delta e^{-\kappa^2/2}}\,,
 \\
\mycorrel^{G}(\pv)
&=&
\frac{ R_l \Delta e^{-\kappa^2/2}}
{ \big(\taut + \kappa^2 +  \Delta e^{-\kappa^2/2}\big)^2}\,.
\label{corr-2}
\end{eqnarray}
\end{subequations}
Note that Eqs.~(\ref{corr}) are structurally identical to their bare counterparts, Eqs.~(\ref{corr-bare}), the only difference being the replacement of the bare reduced temperature $t$ by its renormalized counterpart $\taut$.

The relation~(\ref{relation-tau-t}) between $\taut$ and $t$ determines the existence, or lack thereof, of a transition out of the macroscopically isotropic state. Let us first discuss the high (measurement) temperature regime, in which $\bar{\nop}={\bm 0}$ and the correlators are small.  In this case, the second term on the RHS of  Eq.~(\ref{relation-tau-t}), proportional to $\ldgfour$, can be neglected.  Thus, $\taut$ is essentially the bare (reduced, rescaled) temperature $t$, and the renormalized correlators~(\ref{corr}) reduce to the bare correlators~(\ref{corr-bare}), as expected.

But what happens at lower temperatures? As $t$ is reduced, the correlators become larger, and the corrections due to fluctuations, present in Eq.~(\ref{relation-tau-t}), increase.  The essential question is then the following:  Do the denominators in Eqs.~(\ref{corr}) {\it vanish\/} at some wave-vector $\pv$ for low enough temperature?  If this were indeed to happen, it would signify a continuous transition to some macroscopically symmetry-broken state.  Now, the low-temperature physics of our model depends sensitively on the value of $\Delta$.  For $\Delta < 2$,  the denominator factor in Eqs.~(\ref{corr}) [viz., $\taut + \kappa^2 + \Delta e^{-\kappa^2/2}$] has a minimum at $\kappa = 0$.  Assuming that a continuous transition occurs at some finite temperature ${t}_c$, this denominator would have to vanish for $\kappa = 0$ and $\taut + \Delta = 0$.  For three-dimensional systems, however, the integral of the glassy correlator $\int_p \mycorrel^{G}$ in Eq.~(\ref{relation-tau-t}) has infrared divergence, and thus $\taut({t}_{c})$ diverges, implying that the denominator would diverge, and this contradicts the assumption that this denominator vanishes.  We therefore conclude that in three dimensions and for $\Delta < 2$ there is no continuous transition to any state having long-range nematic order.  For $\Delta > 2$, the factor $\taut + \kappa^2 + \Delta e^{-\kappa^2/2}$ has a minimum at $\kappa^2 = 2\ln (\Delta/2)$.  In this case, to have a continuous transition (towards a macroscopically periodically modulated state) would require the vanishing of the denominator, now at $\kappa^2 = 2\ln (\Delta/2)$.  However, once again, the integral in Eq.~(\ref{relation-tau-t}) is divergent, and therefore no continuous transition occurs.  Hence, owing to network heterogeneities, \textit{the macroscopically isotropic phase is always locally stable, for any positive $\Delta$\/}.
This result agrees with experimental observations, e.g.~by the Urayama group~\cite{urayama},
and the results of simulations performed by Uchida~\cite{uchida-2d, uchida}.

\section{{Correlator structure and low-temperature properties}}
\label{sec;details}

\begin{figure}
	\centering
	\includegraphics[width=0.47\textwidth]{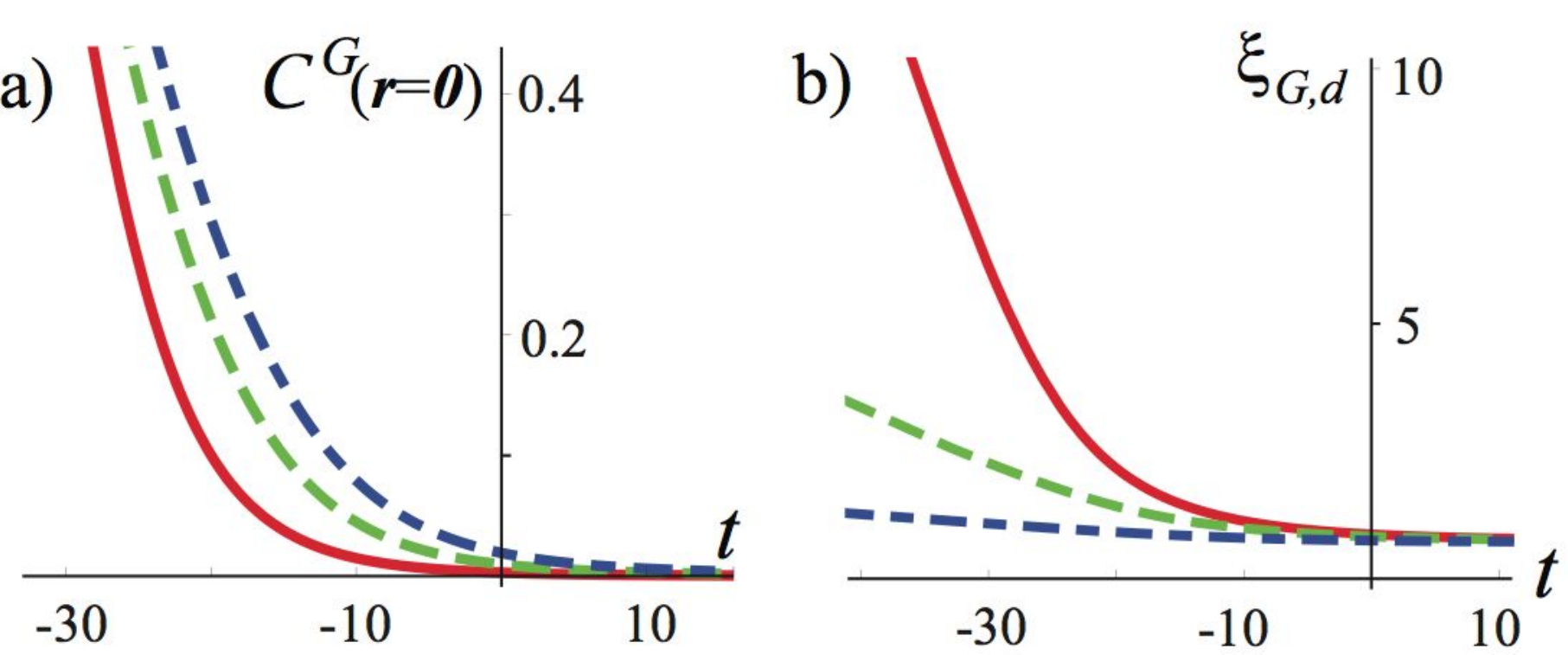}
	\caption{(a)~Glassy order parameter $\mycorrel^{G}(\rv)|_{\rv=\bm{0}}$ (measured in unit $R_l$) and (b)~glassy correlation length $\correllength_{G,d}$ (in unit $\correllength_L$) as a function of $t$ for three cases of weak disorder: $\Delta=0.2$ (red solid), $\Delta=0.7$ (green dashed) and $\Delta=1.5$ (blue dot-dashed), where we have set $w R_l=0.4\pi$.}
\label{fig:inflation1}
\end{figure}

\begin{figure}[b]
	\centering
		\includegraphics[width=0.43\textwidth]{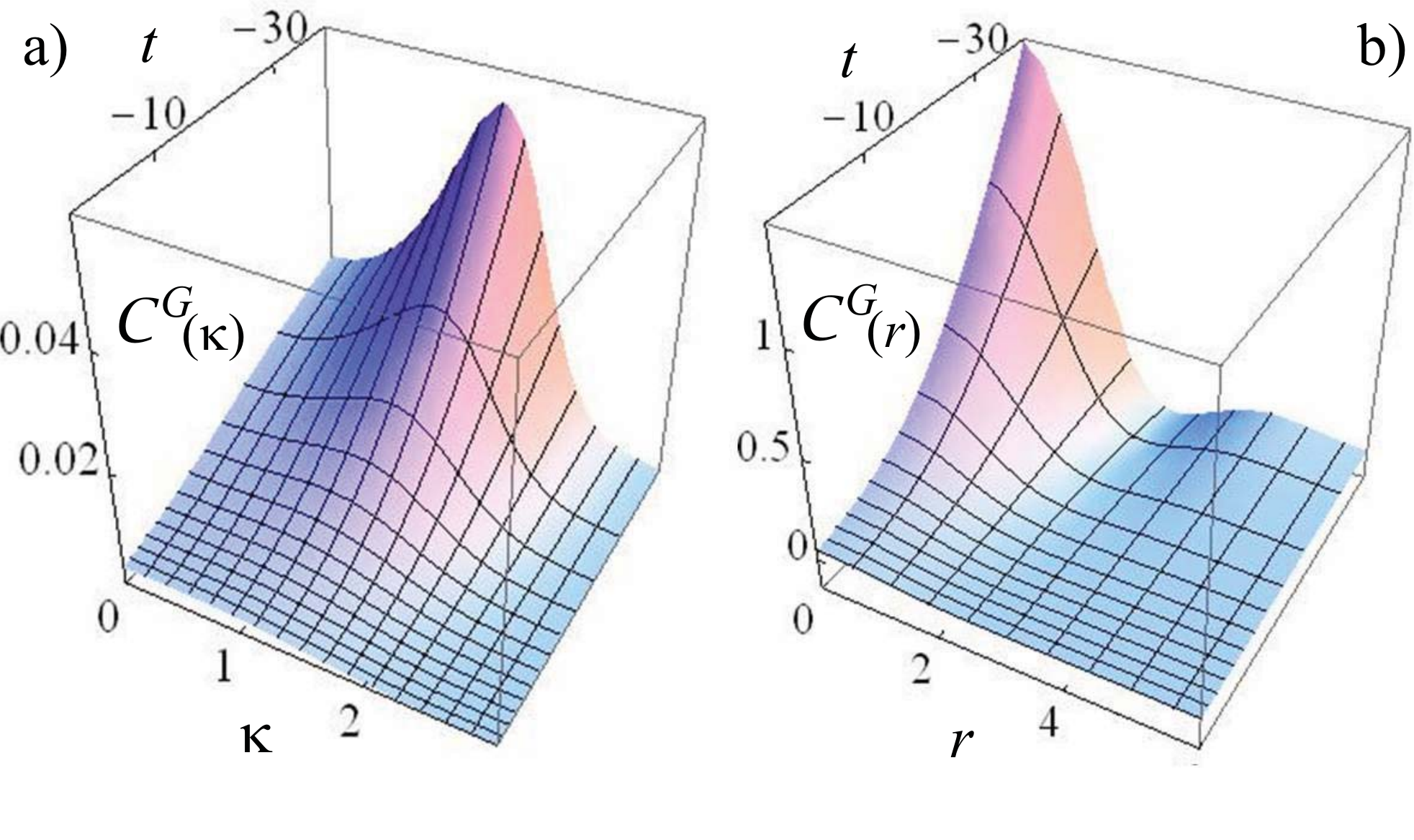}
	\caption{Glassy correlator $\mycorrel^G$ (measured in unit $R_l$) at $\Delta=10$ as a function of reduced temperature $t$ and (a)~reduced wave-vector $\kappa$; (b)~separation distance $\rv$ (in unit $\correllength_L$), with $w R_l=0.4\pi$. Note the progression in~(a) of the peak location from zero wave-vector to nonzero wave-vector with decreasing $t$.  The peak at nonzero wave-vector is associated with the spatial oscillations observed in~(b) for the glassy correlator in real-space.  Note also in~(b) that the glassy order parameter $\mycorrel^{G}(\rv)|_{\rv=0}$ rapidly increases as the temperature is reduced.}
  \label{fig:glasscorr}
\end{figure}

We now unfold the physics encoded in the correlators Eqs.~(\ref{corr}).  First, we look at the glassy order parameter $\mycorrel^{G}(\rv)|_{\rv=\bm{0}}$, which is given by $\int_{p} \mycorrel^{G}(\pv)$.  From Eq.~(\ref{corr-2}), we see that below $t=0$ it increases rapidly with decreasing temperature, becoming, at low temperatures, asymptotically linear in $t$, behaving as ${K |t|}/{7 \ldgfour \xi_L^2}$.  We have computed this order parameter: the results are shown for three cases of weak disorder in Fig.~\ref{fig:inflation1}a and one case of strong disorder in Fig.~\ref{fig:glasscorr}b.

As the fluctuation-response theorem indicates, the static susceptibility $\chi(\pv)$ for local nematic order is related to the thermal correlator via $\chi(\pv) = T^{-1} \mycorrel^{T}(\pv )$. The peak value of $\mycorrel^T$, which for weak disorder (i.e., $\Delta < 2$) occurs at $\kappa^2 = 0$ and for strong disorder (i.e., $\Delta > 2$) occurs at $\kappa^2=2\ln(\Delta/2)$, increases rapidly as the temperature is decreased, as we see from Eq.~(\ref{corr-1}).  Due to the coupling between local nematic alignment and elastic strain, we thus expect there to be a corresponding soft elasticity.  We speculate that this mechanism lies behind the supersoft elasticity of PD IGNEs (cf.~Refs.~\cite{biggins-short,biggins-long}).  A detailed analysis of elasticity of PD IGNEs is however not possible using the current simple model which does not involve the elastic degrees of freedom.  We shall pursue it in a separate publication.


In the weak-disorder regime (i.e. $\Delta < 2$), both correlators have a finite peak at zero wave-vector, indicating spatial decay (but without oscillation).   The characteristic lengthscales $\correllength_{T,d}$ and $\correllength_{G,d}$ over which the  correlators decay can be obtained by examining their small wave-vector behaviours:
\be\label{eq:wkcl}
\correllength_{T,d}^2 \approx \frac{(2-\Delta)}{2(\taut+\Delta)}\correllength_L^2,
\qquad
\correllength_{G,d}^2 \approx 2\correllength_{T,d}^2+\tfrac{1}{2}\correllength_L^2.
\ee
Since $\taut + \Delta$ approaches zero towards low temperature, both lengthscales $\correllength_{G/T,d}$ grow rapidly (see Fig.~\ref{fig:inflation1}b).

\begin{figure}
\centering
		\includegraphics[width=0.47\textwidth]{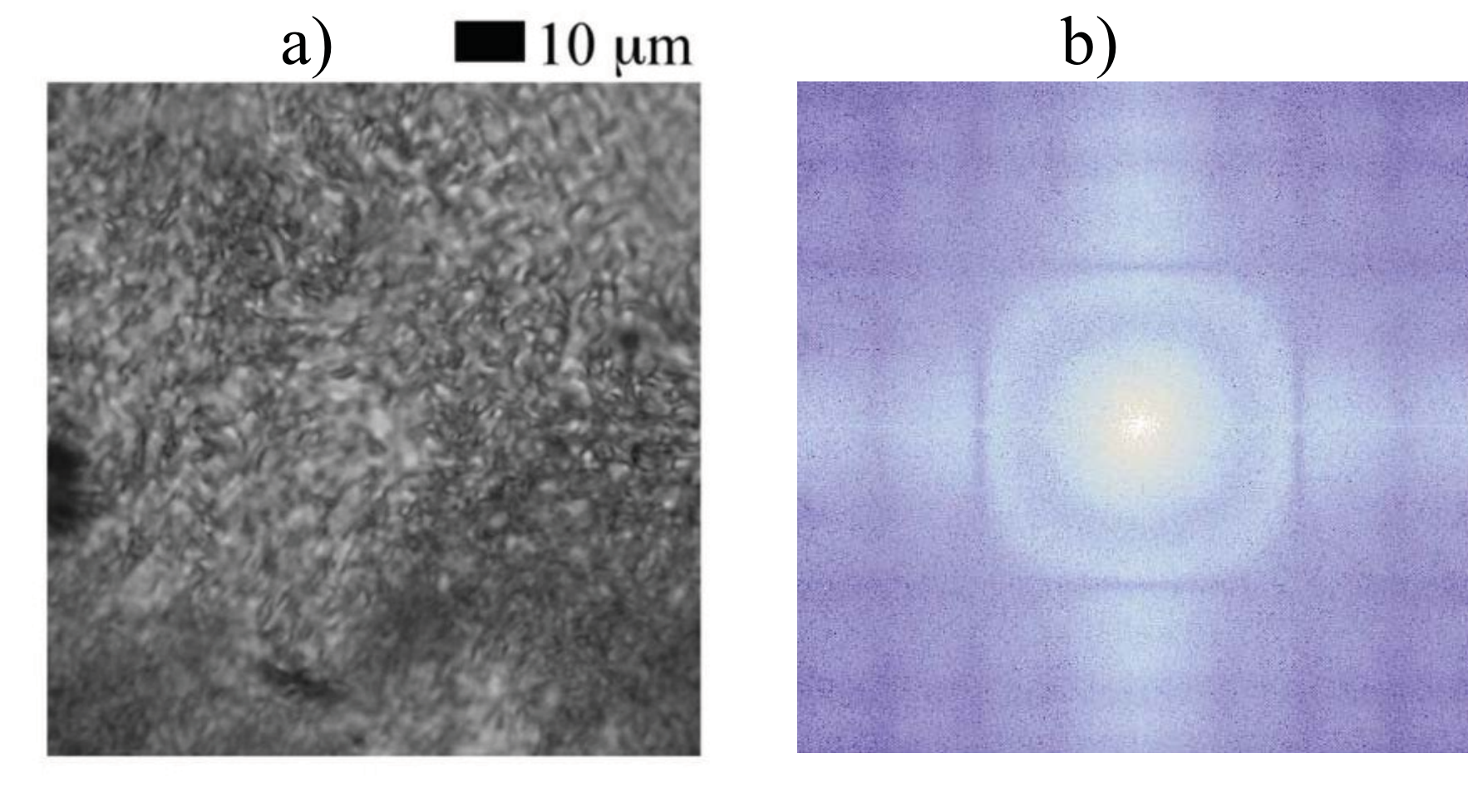}
\caption{(a)~Real-space optical microscopic image of polydomain structure in IGNEs, obtained by the Urayama group. Note the occurrence of a short-range \lq checkerboard\rq\ pattern indicative of oscillatory structure in the local nematic alignment, most evident in the upper-right region of the panel.
(b)~Noise-filtered (Log amplitude-squared) Fourier transformation of the optical image in panel~(a). Note the occurrence of a central high-intensity (bright) peak accompanied by a high-intensity ring, which correponds to the checkerboard pattern in real space.  The shaded rectangular grid is due to the denoising filter. }
\label{fig:urayama}
\end{figure}

\begin{figure}
	\centering
		\includegraphics[width=0.47\textwidth]{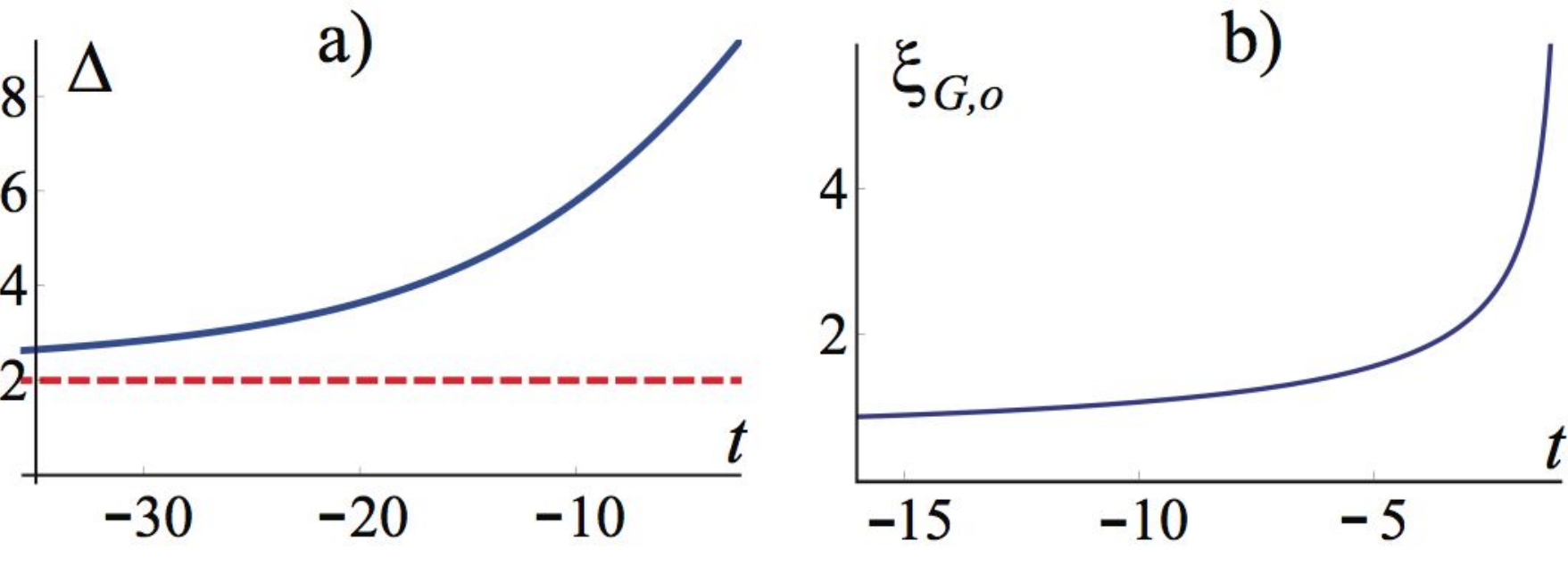}
	\caption{(a)~Crossover diagram for the glassy and thermal correlators, indicating the three
    qualitatively distinct regimes of behavior.
    Above the blue solid line, both correlators oscillate and decay as a function of separation.
    Between the blue solid and red dashed lines, both correlators decay but only the thermal one
    also oscillates.
    Below the red dashed line, both correlators decay but neither oscillates.
    (b)~Oscillation wavelength $\correllength_{G,o}$ (measured in unit $\correllength_L$) as a function of reduced temperature $t$ for $\Delta=10$.
    As $t$ decreases, $\correllength_{G,o}$ eventually saturates at a nonzero value of order $\xi_L$. For both panels, we have set $w R_l=0.4\pi$.}
	\label{fig:phase1}
\end{figure}
In the strong-disorder regime (i.e., $\Delta > 2$), the thermal correlator $\mycorrel^T$ and the glassy correlator $\mycorrel^G$ are each peaked on a spherical shell of wave-vectors: $\mycorrel^{T}(\pv)$ peaks for $|\pv|^2 = 2\ln(\Delta/2)/\correllength_{L}^{2}$, and $\mycorrel^G$ peaks for $|\pv|^2=\correllength_{G,o}^{-2}$, with $\correllength_{G,o}$ obeying
\begin{equation}
\label{eq:nonlinear}
\taut + 4 + (\correllength_L/\correllength_{G,o})^2 -\Delta\thinspace e^{-(\correllength_L/\correllength_{G,o})^2/2}=0.
\end{equation}
We term $\correllength_{G,o}$ the \textit{glassy oscillation length}.  The peaking of $\mycorrel^{T}(\pv)$ and $\mycorrel^{G}(\pv)$ is associated with oscillatory behaviour (i.e., anti-alignment) of the local nematic order, which is superposed on their overall decay in real space.  This echos the polydomain structure with well defined length scale observed by the Urayama group~\cite{urayama}.  Indeed, as shown in Fig.~\ref{fig:urayama}, Fourier transform of their real space image of the PD structure, after denoising and deblurring, shows clearly a ring of local maximum at the scale of about half micron.
%
%
In Fig.~\ref{fig:glasscorr}, we show examples of how the shell radius for the glassy correlator progresses from the origin to a nonzero wave-vector magnitude and, furthermore, how the oscillatory decay develops in real space, as the measurement temperature is decreased.  The crossover, from non-oscillatory to oscillatory decay as the disorder strength is increased, is shown in Fig.~\ref{fig:phase1}a.  It can be deduced from Eq.~(\ref{eq:nonlinear}) that, as the temperature is decreased, $\correllength_{G,o}$ decreases and ultimately saturates at the temperature-independent \textit{thermal oscillation length} $\correllength_{T,o}$ ($=\correllength_{L}/\sqrt{2\ln(\Delta/2)}$).  Provided that $\correllength_{G,d}$ exceeds $\correllength_{G,o}$, the length $\correllength_{G,o}$ will reflect the size of the locally aligned nematic domains; therefore, the decrease of $\correllength_{G,o}$ (shown in Fig.~\ref{fig:phase1}b) will reflect the shrinking of the domain size with decreasing temperature.


\section{Conclusion and outlook}

Via a strategy rooted in vulcanization theory, we have developed a unified approach to the statistical physics of nematic elastomers, in which nematic ordering---both in the preparation and the measurement states---and network heterogeneity are naturally incorporated.  It is noteworthy that the quenched disorder naturally emerges from this approach differs in essential ways from the traditional quenched random field approach.  This difference evinces a strength of vulcanization theory.  Using a variational method, we show that the macroscopically isotropic state of isotropic-genesis nematic elastomers is always thermodynamically stable and, furthermore, that for sufficiently high disorder strengths, short-range oscillatory structure of the local nematic alignment arises.

%

Apart from its relevance to the specific subject of liquid crystalline elastomers, the present work brings to light a more general issue, viz., that the concept of a quenched random field  should be broadened to incorporate not only the traditional, \lq frozen\rq\ type, which does not fluctuate thermally, but also the type necessary for understanding media such as liquid crystalline elastomers, in which the frozen nature of the random field is present only at longer lengthscales, fading out as the lengthscale progresses through a characteristic localization length, owing to the thermal position fluctuations of the network's constituents.  The framework elucidated in the present work can be extended, with suitable modifications, to explore the statistical physics of other randomly crosslinked systems, such as smectic elastomers and various biological materials.

\begin{acknowledgments}
We thank Kenji Urayama for informative discussions on experimental results and Xiaoqun Zhang for valuable assistance with image processing.  This work was supported by the National Science Foundation via grants~DMR06-05860 and DMR09-06780, the Institute for Condensed Matter Theory and the Frederick Seitz Materials Research Laboratory at the University of Illinois at Urbana-Champaign, the Institute for Complex Adaptive Matter, and Shanghai Jiao Tong University.
\end{acknowledgments}



\begin{thebibliography}{99}
\bibitem{footnote-one-replica}
For $\alpha = \beta$, we can integrate out all $\xv^{\gamma}$ for $\gamma \neq \alpha$.   The $\nabla^{\alpha} \nabla^{\alpha} \bar{\Omega} \nabla^{\alpha} \nabla^{\alpha} \bar{\Omega}$ part in Eq.~(\ref{coupling}) then reduces to a function of $\xv^{\alpha}$ only.  Due to the invariance of the saddle point order parameter $\bar{\Omega}$ under simultaneous translation of $\hat{r}:= (\rv^0,\ldots, \rv^{n})$, however, this function must be a trivial constant.
\bibitem{footnote:shearmodulus}
At the mean field level, the shear modulus of a network is proportional to $G^2 \xi_L^{-2}$, and so is the disorder strength $\Delta$.
\bibitem{footnote:kernel}
Strictly speaking, the structure of the kernel $\Gamma^{\alpha\beta}$ should allow for anisotropy for the case of $\bar{\nop}\neq 0$.  Our main analysis, however, only concerns the isotropic state, for which $\bar{\nop}=0$.
\bibitem{footnote:rsb}
According to the standard understanding, a replica symmetry breaking at low temperature implies that the system freezes into one of many nearly degenerated states.  Such a system should exhibit irreversible glassy properties such as hysteresis in stress-strain curve.  These properties have not been observed in polydomain nematic elastomers.
\end{thebibliography}
\end{document}